\documentclass[useAMS,usenatbib]{mn2e}

\usepackage {graphicx}
\usepackage {times}

\title[A QSO host galaxy and its Ly$\alpha$ emission at z=6.43 ]{A QSO host galaxy and its Ly$\alpha$ emission at z=6.43\footnotemark[0]\thanks{Based  on data collected at Subaru Telescope, which is operated by the National Astronomical Observatory of Japan.}}
\author[Goto et al.]{Tomotsugu Goto$^{1,2}$ %\footnotemark[1]
 \thanks{E-mail:tomo@ifa.hawaii.edu} \thanks{JSPS SPD Fellow}
 , Yousuke Utsumi$^{2,3}$,
Hisanori Furusawa$^{2}$,
\newauthor    Satoshi Miyazaki$^{2,3}$,
 and 
Yutaka Komiyama$^{2}$
\\
$^{1}$
Institute for Astronomy, University of Hawaii
2680 Woodlawn Drive, Honolulu, HI, 96822, USA\\
$^{2}$National Astronomical Observatory, 2-21-1 Osawa, Mitaka, Tokyo
181-8588,Japan\\
$^{3}$ Department of Astronomical Science,The Graduate University for Advanced Studies,
 2-21-1 Osawa, Mitaka, Tokyo 181-8588, Japan\\\\
\\
}

\begin{document}
%--- DRAFTCOPY: COMMENT OUT IF NOT NEEDED -------
% Prints a large "DRAFT" diagonally across each page
% Does not show up in TeXview
% \typeout{Prints "DRAFT" on each page; does not show in TeXView}
% \special{!userdict begin /bop-hook{gsave 200 30 translate
% 65 rotate /Times-Roman findfont 216 scalefont setfont
% 0 0 moveto 0.90 setgray (DRAFT) show grestore}def end}
%%------------------------------------------------
\def\Hg{H$\gamma$}
\def\Hd{H$\delta$}

\date{Accepted 2009 August 04. Received 2009 June 03; in original form 2009 February 24}
%\date{\today; in original form 2009 Feb 23}

\pagerange{\pageref{firstpage}--\pageref{lastpage}} \pubyear{2009}

\maketitle

\label{firstpage}

\begin{abstract} 
Host galaxies of highest redshift QSOs are of interest; they provide us with a valuable opportunity to investigate physics relevant to the starburst-AGN connection at the earliest epoch of the Universe, with the most luminous black holes.
 
 Here we report an optical detection of an extended structure around a QSO at z=6.43 in deep $z'$ and $z_r$-band images of the Subaru/Suprime-Cam. Our target is CFHQSJ2329-0301 (z=6.43), the highest redshift QSO currently known.
We have carefully subtracted a PSF constructed using nearby stars from the images.
 After the PSF (QSO) subtraction, a structure in the $z'$-band extends more than 4'' on the sky ($R_e$=11 kpc), and thus, is well-resolved (16$\sigma$ detection). The PSF-subtracted $z_r$-band structure is in a similar shape to that in the $z'$-band, but less significant with a 3 $\sigma$ detection.
In the $z'$-band,  a radial profile of the QSO+host shows a clear excess over that of the averaged PSF in 0.8-3'' radius. 
 
 Since the $z'$-band includes a Ly$\alpha$ emission at z=6.43, we suggest the $z'$ flux is a mixture of the host (continuum light) and its Ly$\alpha$ emission, whereas the $z_r$-band flux is from the host.
 Through a SED modeling, we estimate 40\% of the PSF-subtracted $z'$-band light is from the host (continuum) and 60\% is from Ly$\alpha$ emission.
 The absolute magnitude of the host is $M_{1450}$=-23.9 (c.f.  $M_{1450}$=-26.4 for the QSO). 
 A lower limit of the SFR(Ly$\alpha$) is 1.6 $M_{\odot}$ yr$^{-1}$ with stellar mass ranging  6.2$\times 10^8$ to 1.1$\times 10^{10} M_{\odot}$ when 100 Myrs of age is assumed. 
 The detection shows that a luminous QSO is already harbored by a large, star-forming galaxy  in the early Universe only after $\sim$840 Myr after the big bang.
 The host may be a forming giant galaxy, co-evolving with a super massive black hole. 
\end{abstract}


\section{Introduction}

%\begin{itemize}
% \item SB-AGN connection
% \item Margorian relation
%\end{itemize}

%  It has been found that many nearby galaxies host central active galactic nuclei (AGN).
 There have been accumulating evidence that galaxies and active galactic nuclei (AGN) co-exist.
  Evidence of starburst features have been detected in numerous AGNs
     \citep{1997ApJ...485..552M,2005MNRAS.356..270C}.
 Half of the  ultraluminous infrared galaxies contain simultaneously an AGN and  starburst activity \citep{1998ApJ...498..579G,2005MNRAS.360..322G}.
 Post-starburst signatures have been found in AGNs \citep{1999ApJ...520L..87B,2000MNRAS.318..309D,2006MNRAS.369.1765G}.  
 The discovery of tight correlation between black hole mass and bulge velocity dispersion provides another evidence that the formation of bulge and the central black hole may be closely linked \citep{1998AJ....115.2285M,2000ApJ...539L...9F,2000ApJ...539L..13G,2001MNRAS.324..757G,2004ApJ...613..109H,2004ApJ...615..645O}.%  (Onken et al. 2004).
%It has been known there is a close correlation between the central black hole mass and the properties of the spheroidal stellar population that surrounds it (Onken et al. 2004; Ferrarese \& Merritt 2000). 
The correlation implied the co-evolution of a super-massive black hole and its host galaxy. 
%Often a starburst and an AGN are found in the same galaxy (the so-called starburst-AGN connection).
 To understand galaxy formation theory and the AGN mechanisms, it is important to characterize a possible relationship between the galaxy formation and AGNs. 

 At high luminosity end of the AGN, QSOs are not exception;
At low redshift,  QSOs are predominantly hosted by luminous, massive, bulge dominated galaxies 
\citep{2002ApJ...576...61H,2003MNRAS.340.1095D,2003ApJ...596..830P,2004MNRAS.355..196F}, irrespective of radio power.
%Recently, a spectroscopic study of QSO host galaxies at $z\sim0.3$ was performed by Letawe et al. (2007). Most host galaxies were found to have Sc-like populations, showing large amounts of ionised gas and signs of interaction.
 At high redshift, evolution of the QSO host has been reported;
 \citet{2008A&A...478..311S}  found hosts with stellar populations indicating recent massive star formation at $z\approx3$ . 
 \citet{1995MNRAS.275L..27A,1998MNRAS.296..643A} found SFR of $>$100-200M$_{\odot} yr^{-1}$ for z$\sim$2 QSO hosts.
 \citet{2008A&A...488..133V} also found moderate rate of SFR for QSO hosts at z=0.87 and 2.85.
\citet{2002AJ....123.2936H,2006AJ....131..680H} found that high redshift QSO host galaxies are irregular and highly disturbed.
 \citet{2005PASP..117.1250H} reported a significant young stellar population with a possible spiral structure in two hosts at z=5.
 However also note that some luminous QSOs are reported to reside in luminous massive early-type galaxies at $1<z<2$ \citep{2007ApJ...660.1039K}, and at $2<z<3$ \citep{2008ApJ...673..694F}. 
% On the other hand, 
 %The different results seem to indicate that stellar populations of QSO host galaxies depend on the luminosity of the QSO itself, the radio-loudness and the redshift.
% The luminosity of host galaxies seems to increase with redshift, with 
 The host galaxies of radio-quiet QSOs are on average less luminous than the ones of radio-loud QSOs \citep[e.g.,][]{2001MNRAS.326.1533K,2007ApJ...660.1039K}.
 The apparent co-evolution of QSO and host galaxies leads to the question if QSO activity and star-formation are caused by the same physical mechanism, or one triggers the other.

 Although a proper investigation of high-z QSO hosts would shed light on the QSO-galaxy co-evolution, QSO host galaxies at highest redshifts of $z>6$ have not yet been studied extensively, since due to high QSO-to-host light ratios, extremely deep, high resolution images are required to assure a sufficiently high signal-to-noise ratio.
 A pioneering attempt by \citet{2005PASP..117.1250H} did not resolve host galaxies of QSOs at z=6.23 and z=6.35.

In this paper, we investigate a host galaxy of a QSO at z=6.43 using deep optical images taken with the Suprime-Cam/Subaru equipped with new red-sensitive CCDs. This is the highest redshift QSO as of the date of writing, and thus, provides us with an unprecedented opportunity to shed light on the co-evolution of galaxy and QSO at the massive end of super massive black holes with $\sim10^9M_{\odot}$, at the earliest epoch when the age of the Universe was only $\sim$840 Myrs old. 

  Unless otherwise stated, we adopt the WMAP cosmology: $(h,\Omega_m,\Omega_L) = (0.7,0.3,0.7)$ \citep{2008arXiv0803.0547K}.

% It has been observationally known that there is a tight correlation between the bulge luminosity and the mass of the central black holes observed in nearby inactive elliptical galaxies (Ferrarese 2006).
%
%the correlation between the super massive black holes and galaxy evolution.

\begin{figure}
\begin{center}
\includegraphics[scale=0.4]{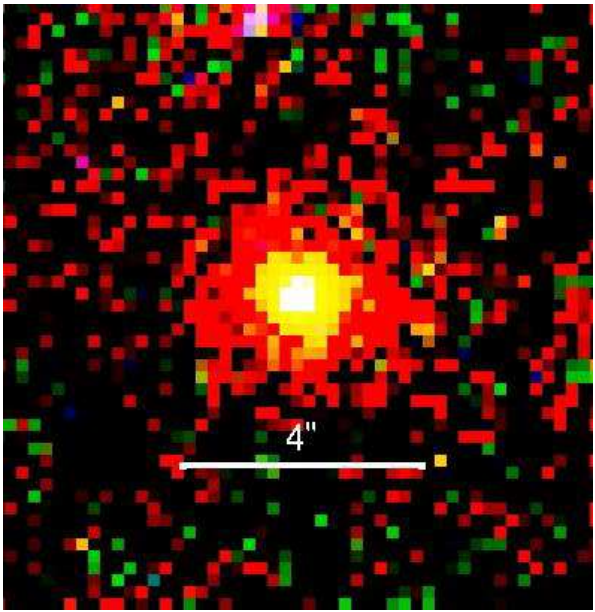} 
\end{center}
\caption{
Composite pseudo-color image of the QSO (CFHQSJ2329-0301). The RGB colors are assigned to $z', z_r$ and $i'$-bands, respectively.
The figures are north up, east  left.
}\label{fig:composite}
\end{figure}

%\begin{figure*}
%\begin{center}
%%\includegraphics[scale=0.9]{/home/tomo/work/galfit-example/utsumi_QSO/081209_galfit_out_z3psf.ps} %z3psf
%%\includegraphics[scale=0.9]{/home/tomo/work/galfit-example/utsumi_QSO/081210_galfit_out_z5psf.ps} %z5psf
%\includegraphics[scale=0.9]{/home/tomo/work/galfit-example/utsumi_QSO/090210_hostimage2.ps} %z5psf
%\end{center}
%\caption{
%(left) The $z'$-band image.
%(middle) A PSF constracted using nearby stars.
%(right) A PSF subtracted image. 
%}\label{fig:z_subtraction}
%\end{figure*}

\begin{figure*}
\begin{center}
\includegraphics[scale=0.87]{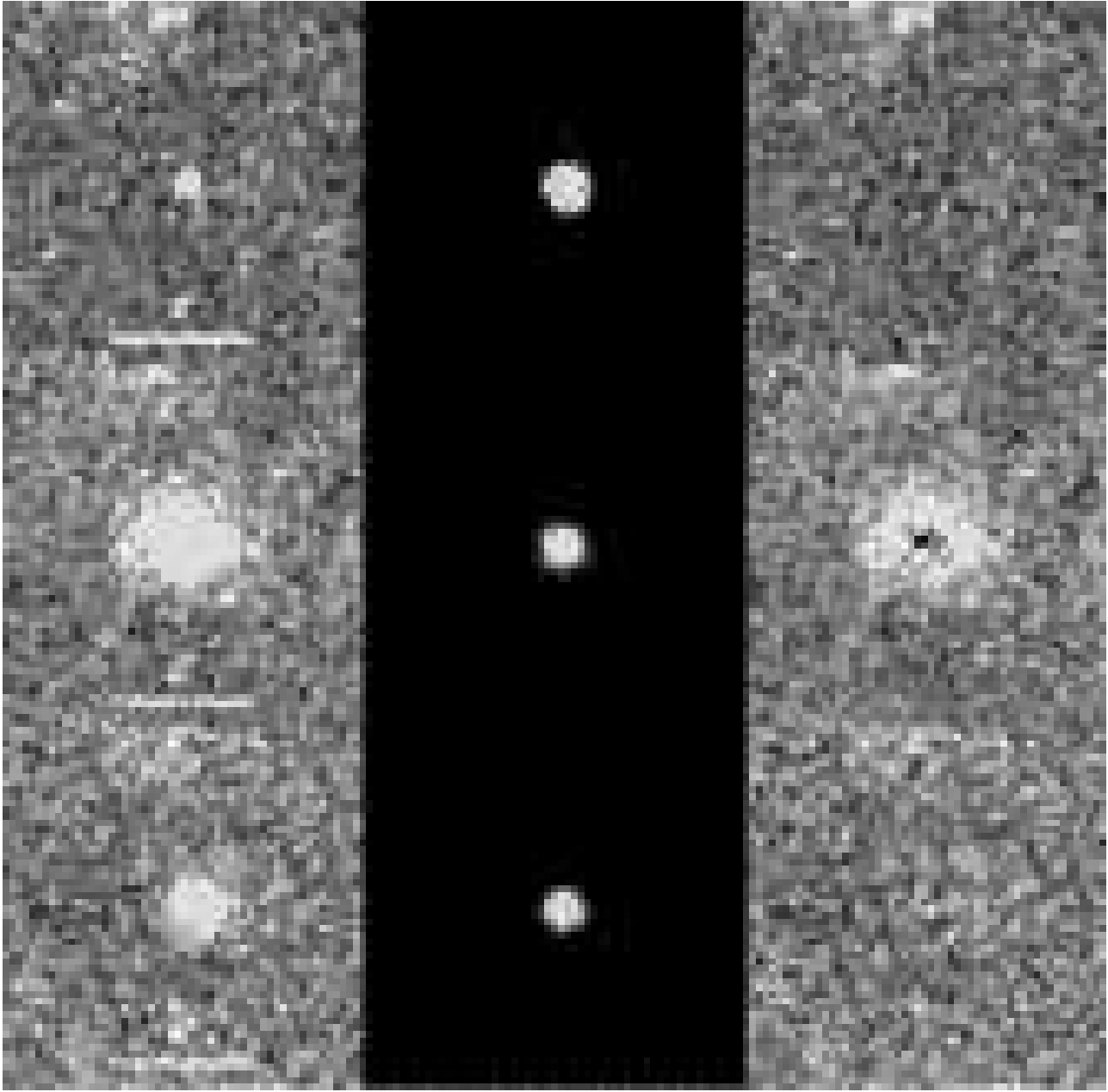}
\end{center}
\caption{
The upper, middle, lower panels are for $i'$, $z'$ and $z_r$-band, respectively.In each line, 
the left panels are reduced images. 
The middle panels are PSFs constructed using nearby stars.
The right panels show  residuals from the PSF subtraction. 
All figures are north up, east  left.
}\label{fig:all_subtraction}
\end{figure*}

\begin{table*}
% \centering
 \begin{minipage}{180mm}
  \caption{Target information adopted from \citet{2007AJ....134.2435W}}\label{tab:targets}
  \begin{tabular}{@{}llllrrcccc@{}}
  \hline
   Object & RA & DEC & z& $i'_{AB}$ & $z'_{AB}$ & $J$ & $M_{1450}$  \\ 
 \hline
 \hline
CFHQS J232908-030158 &23:29:08.28  & -03:01:58.8 & 6.43 & $>$25.08 &	21.76$\pm$0.05&	21.56$\pm$0.25& -26.4\\
\hline
\end{tabular}
\end{minipage}
\end{table*}

\begin{table}
 \centering
 \begin{minipage}{80mm}
  \caption{Basic properties of the constructed PSFs. The pixel scale is (0.2''/pix). The FWHM is obtained through a Gaussian fit.}\label{tab:PSFs}
  \begin{tabular}{@{}llllrrcccc@{}}
  \hline
   Filter  & FWHM (pix) & ellipticity & \\% 3$\sigma$ limit mag (26 pix aperture) \\
 \hline
 \hline
$i'$ ($\lambda_e$=768nm)  & 2.45 (=0.49'') & 0.04 & \\
$z'$ ($\lambda_e$=911nm)  & 2.25 (=0.45'') & 0.06 & \\
$z_r$ ($\lambda_e$=988nm) & 2.69 (=0.54'') & 0.07 & \\
\hline
\end{tabular}
\end{minipage}
\end{table}

\begin{table*}
% \centering
 \begin{minipage}{180mm}
  \caption{Magnitudes and results of the fit. 
%Note that 1 $\sigma$ upper limit of the host magnitudes are in a 26 pix diameter aperture.
}\label{tab:fit}
  \begin{tabular}{@{}llllrrcccc@{}}
  \hline
%   Object     &            & \multicolumn{4}{c}{Flux density (Jy)%}
%   Object & $i'_{AB}$ & $z'_{AB}$ & $zr_{AB}$ & $R_e (arcsec)$ & axis ratio (b/a) & n(exponent)  \\ %(deVauc=4, expdisk=1)  
   Object & $i'_{AB}$ & $z'_{AB}$ & ${z_r}_{AB}$   \\ %(deVauc=4, expdisk=1)  
 \hline
 \hline
%galfit
% QSO & &21.5& &  0.5 & 0  & ---\\
% Host & &23.7& & 2.0   & 0.74 & 0.0190\\
%\hline
%imexam  
% QSO &  &21.39& &  --- & 0  & ---\\
% Host & &23.55&  & ---   & --- & --- & 16\\
%\hline
%GAIA
 QSO+host &  25.54$\pm$0.02    & 21.165 $\pm$ 0.003 & 21.683$\pm$0.007  \\
% Host &   $>$26.94    &23.63 $\pm$ 0.01& $>25.69$  & ---   & --- & --- & 16\\
% Host &   $>$25.66 (3$\sigma$ limit)    &23.63 $\pm$ 0.01& $>$24.41 (3$\sigma$ limit)  & 2.0   & 0.74 & 0.019 & 16\\
%Host &   $>$25.34 (1$\sigma$ limit)    & 23.50 $\pm$0.10 & $>$24.22 (1$\sigma$ limit)  & 2.0   & 0.74 & 0.019 & 16\\
%Host &   $>$25.34 (1$\sigma$ limit)    & 23.5$\pm$0.3 (16$\sigma$) & 24.3$\pm$0.2 (3$\sigma$)& 2.0$\pm$0.1   & 0.74$\pm$0.03  & 0.019 \\
Host &   $>$25.34 (1$\sigma$ limit)    & 23.5$\pm$0.3 (16$\sigma$) & 24.3$\pm$0.2 (3$\sigma$)\\
1$\sigma$ sky &   25.44 (26 pix diameter)    & 24.90 (26 pix diameter)   & 25.46 (18 pix diameter)    \\
\hline
\end{tabular}
\end{minipage}
\end{table*}

\section{Observation}
\label{Observation}

%RA : 23:29:08.28
%DEC : -03:01:58.8
%
%CFHQS J232908-030158...	23 29 08.28	-03 01 58.8	>25.08a	21.76 ± 0.05	21.56 ± 0.25	>3.32	0.18 ± 0.25
%
%CFHQS J2329-0301...	6.43	Gemini	2006 Nov 26	1300	1.0	3600	0.7	-25.23

Our target is CFHQSJ2329-0301, which is a QSO at z=6.43 \citep{2007AJ....134.2435W}. Characteristics of the target is summarized in Table \ref{tab:targets}. Assuming Eddington luminosity, its black hole mass is estimated to be 1.0$\times 10^9 M_{\odot}$ \citep{2001AJ....122.2833F}.

 Observations were carried out on August 1, 26, and 27, 2008 with the Suprime-Cam \citep{2002PASJ...54..833M}, on which new red sensitive CCDs have been just installed (Utsumi et al. in prep.). 
 The sensitivity of the new CCDs is 1.3 times better in $z'$-band, and 1.9 times better in $z_r$-band ($\lambda_{effective}$=988 nm). 
 The data were originally taken to investigate the QSO environment, which will be presented elsewhere (Utsumi et al. in prep.).
 Since the observation was partly for an engineering purpose to test the new CCDs, we used various exposure times with  360s $\times$ 10 in $i'$,  100sec $\times$ 7, 300sec $\times$ 2, 400sec $\times$ 3, 500sec $\times$ 6, 700sec $\times$ 2 in $z'$ and  500sec $\times$ 19 in $z_r$. The dithering pattarns were mostly circular with separations of 0.5-2 arcmins.
 The total exposure time on the target was 3600 sec in $i'$, 6900 sec in $z'$, 9500 sec in $z_r$-bands.
 We used a nearby field observed both by the SDSS and ourselves for a photometric calibration.
 A slight difference between the SDSS and the Subaru filters are compensated using the \citet{1983ApJS...52..121G} stellar spectral library.
 The data are reduced using the modified version of the pipeline. 
 The details of the reduction is described in Utsumi et al. (in prep.). 
 The 2$\sigma$ limiting magnitudes are  $i'$=26.73, $z'$=25.79, and $z_r$=25.09 ABmag within 2'' aperture. 
 Here, 1$\sigma$ sky magnitudes are computed by randomly placing a 2''aperture in the blank (sky) part of the image \citep{2002AJ....123...66Y,2004ApJ...611..660O}.
%26.2, 23.3 mag, and 25.5 mag (80\% complete xxxx ) for a point source in  $z'$, $z_r$, and $i'$-bands, respectively. 

 The pixel scale of the Suprime-Cam is 0.2''/pixel. The seeing size of the images are $\sim$0.5 arcsec (see Table\ref{tab:PSFs}).
 An  $i',z$ and $z_r$ composite image of the QSO is shown in Fig.\ref{fig:composite}, where the RGB colors are assigned to $z', z_r$ and $i'$-bands, respectively.
 Without even subtracting the central PSF, it is immediately noted that the QSO has an extended structure in the $z'$-band (red in the composite). 

% In the left panels of Fig.\ref{fig:ip_subtraction}, we show the QSO images in $i',z$ and $z_r$-bands, separately. Although the QSO looks like a point source in $i'$ and $z_r$, it is immediately noted that the QSO has an extended structure in the $z'$-band image. 
% As we investigate in the next section, the structure extends $\sim$4'' and thus already obvious in visual inspection compared to $\sim$0.5'' of seeing size.

\section{Results}\label{results}

\subsection{PSF subtraction}\label{sec:psf}

 In this section, we subtract a PSF from the QSO image to investigate a possibly extended structure around the QSO. 
 In this process, it is critical to carefully construct the PSF model using the observed data since at such a high redshift the PSF has to be known very precise to be able to see a marginally extended host galaxy. 
 Using observed stars instead of analytical functions is especially important for $z_r$-band where 0.5\% of the flux is known to bleed in horizontal spikes due to the filter characteristics.
  In each band, we have carefully picked $\sim$20 nearby stars within 200'' in the mosaic image to avoid a possible PSF variation across the image. 
 These stars are picked to be relatively isolated, and ambiguous stars according to their radial profile.
Stars blended with a nearby object are excluded from the selection.  
We have specifically picked stars as bright as, but not brighter than the QSO itself to avoid a possible non-linearity effect at the central pixels of bright stars. %正しいか？
 We have checked a choice of a different set of PSF stars does not affect results.
Then we used {\ttfamily  DAOPHOT} in {\ttfamily  iraf} to construct an average PSF. The constructed PSF models in each band are shown in the middle panels of Fig.\ref{fig:all_subtraction}.
Basic properties of the PSF in each band are presented in Table \ref{tab:PSFs}.

 Then we use {\ttfamily  galfit} \citep{2002AJ....124..266P}  to subtract the constructed PSFs from the QSO images shown in the left panels of Fig.\ref{fig:all_subtraction}.  
 At first, {\ttfamily  galfit}  was set to fit only the PSF and any remaining sky. The residuals are shown in the right panels of Fig.\ref{fig:all_subtraction}. 
 We discuss results from each band in the following subsections. 
  
\subsubsection{$i'$-band}
 In the $i'$ band (the top panels), no flux is left on the residual image, i.e., the QSO is consistent with a PSF in the $i'$ band. 
 This is not surprising since bluewards of the Lyman break, little light from the host galaxy can escape from the heavy absorption by the neutral hydrogen.
% As a consistency check, we also subtract the PSF from the observed stars around QSO in Appendix \ref{sec:PSF-PSF}. 

\subsubsection{$z'$-band}
 In the $z'$-band shown in the middle panels of the Fig.\ref{fig:all_subtraction}, there clearly remains an extended structure, subtending 4'' after the PSF subtraction. There is a little over-subtraction at the very centre, making the structure looks like a dough-nut shape perhaps because the PSF fit is affected by the extended structure. We have applied the same PSF subtraction to nearby stars for consistency check in Appendix  \ref{sec:PSF-PSF}. None of the PSF fits to stars yielded such an extended structure in Appendix \ref{sec:PSF-PSF}. 
  We performed an aperture photometry using a 13 pix (=2.6'') of radius on the structure. The measured magnitude is $z'$=23.5$\pm$0.3.
 We have computed a 1-$\sigma$ of the sky by randomly placing the same size aperture on the blank regions of the image. The 1 $\sigma$ sky noise is estimated to be 24.90 in $z'$ band.
Compared to the sky noise, the detection is 16$\sigma$, i.e., the structure is well-resolved, and detected. 
% Therefore, we suggest this is a detection of the QSO host galaxy at z=6.43.

 In addition to the PSF subtraction, 
 in the left panel of Fig.\ref{fig:radial_profile}, we show $z'$-band radial profiles of QSO+host in the blue solid line, and of the constructed PSF in the red dashed line. The profiles are normalized to be unity at a maximum.
 The solid, dashed, and short-dashed lines connect median in each sample. 
 The error bars denote the errors in determining the median at each radii.
 %, assuming a possible pixel-to-pixel correlation of a conservative three pixel scale during the image transformation. 
 At above the radius of 4 pixel ($>$0.8''), the figure shows that the QSO+host is clearly more extended than the PSF, showing the extended structure is well resolved.

We note that  there may be a point-source like structure to the right of the residual image in the $z'$-band in the middle-right panel of Fig.\ref{fig:all_subtraction}. However, this is not likely a contamination from a foreground star since the possible point-source is not detected in $i'$-band, and thus quite red. Instead, it is more likely to be a star-forming region/substructure associated with the QSO.

\subsubsection{$z_r$-band}
 In the $z_r$ band image in the bottom left panel of Fig.\ref{fig:all_subtraction}, there may be a possible elongation from north-east to south-west, in a similar orientation to the extended structure in the $z'$-band.
 Although the residual image on the bottom-right panel is much less obvious than in the $z'$-band, there is a hint of remaining flux.  We performed an aperture photometry with an optimal 18 pixel diameter, obtaining 24.3$\pm$0.2 mag of flux. 
 The 1$\sigma$ sky noise measured by randomly placing the same size apertures on the blank regions of the image is 25.46 mag, i.e., the extended structure is at a 3 $\sigma$ level. The errors on the PSF subtraction is 0.2 mag, and thus smaller.
 We also run the source extractor on the PSF subtracted image, detecting the host at a 2.6 $\sigma$ level over the local background.
 
\begin{figure*}
\begin{center}
\includegraphics[scale=0.35]{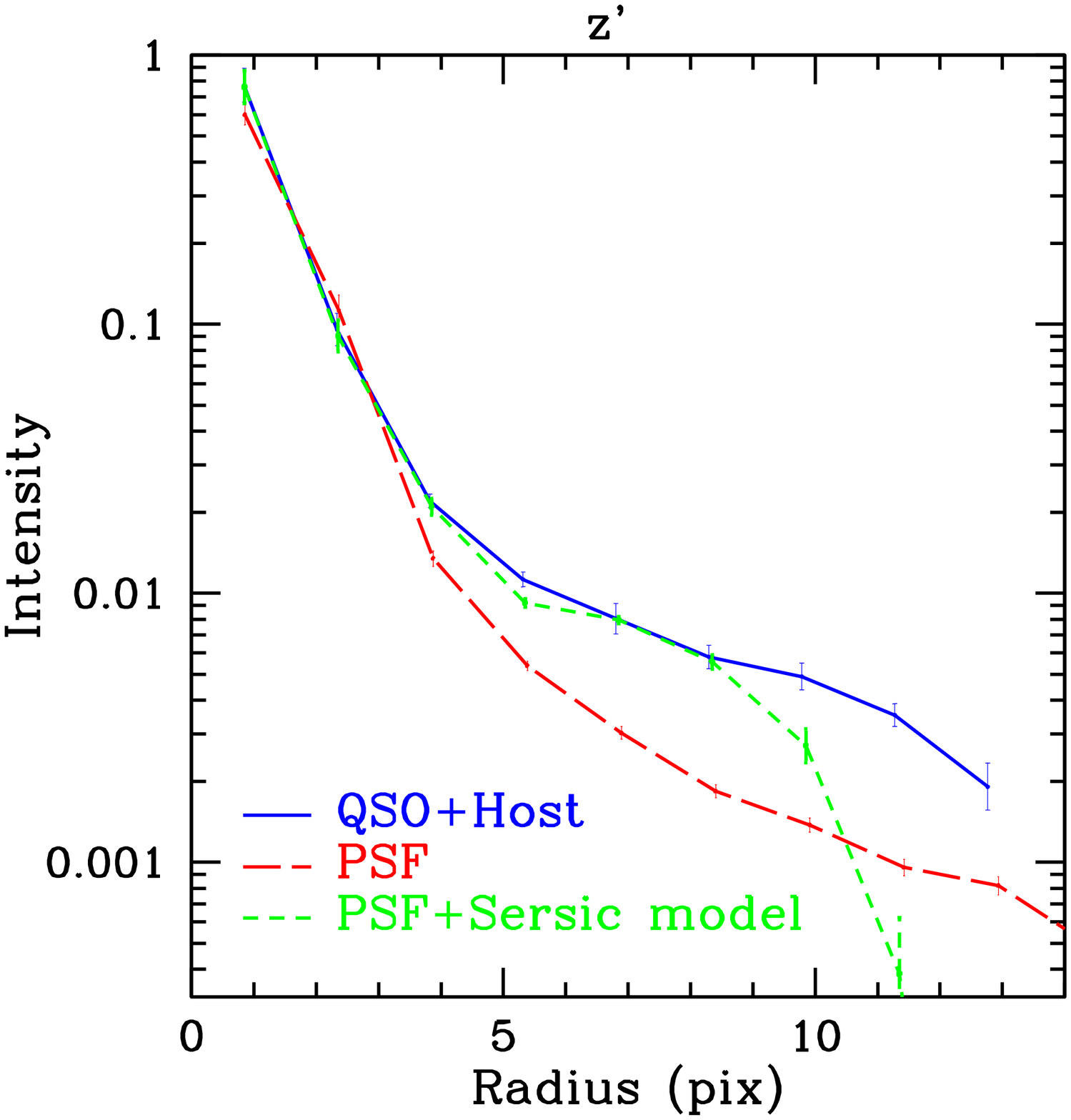}
\includegraphics[scale=0.35]{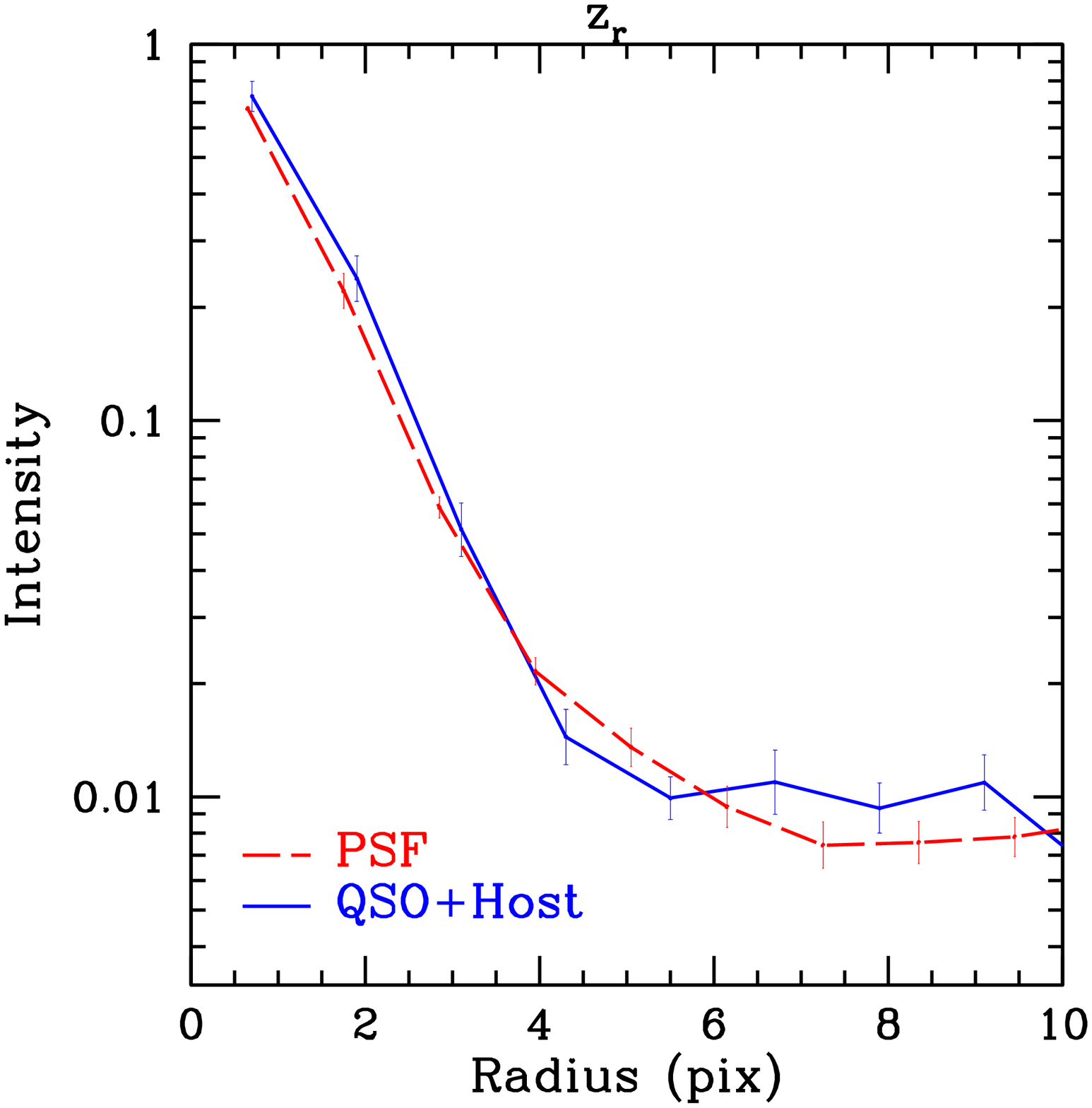}
\end{center}
\caption{
Radial profiles of QSO+host (blue solid line), the constructed PSF (red dashed line), and the PSF+S\'ersic model (green short-dashed line) in the $z'$-band (left). The right panel is for the $z_r$-band. %The solid, dashed, and short-dashed lines connect medians. 
Profiles are normalized at a maximum value.
The pixel scale is 0.2''/pix.
}\label{fig:radial_profile}
\end{figure*}

 The radial profiles in the $z_r$ are shown in the right panel of Fig.\ref{fig:radial_profile}. Although the extended structure is less obvious, the QSO+host has a 1$\sigma$ excess over the PSF in three consecutive radii at 6-9 pixel. 

 The extended structure becomes more visible in a 10-pix box-car smoothed image in Fig.\ref{fig:zr_smooth}.
  For a sanity check, we fit the PSF to nearby stars in Appendix \ref{sec:PSF-PSF}, where we do not see any clear excess in the residuals even after box-car smoothed.

 %After the psf subtraction,  no significant extension was found, but
%there is a little remaining flux at the centre of the image (the bottom panels of Fig.\ref{fig:ip_subtraction}). However, these are likely to be a leftover from the pixelization problem, i.e., due to the limited size of the pixels, there often remains under/over-subtraction at the centre depending what part of the pixel the actual centre of the object falls on, and thus, the errors dominate the host's signal at the centre of the image. In the literature, the pixelization problem can often be found in a PSF subtraction \citep{2006ApJS..166...89G,2007ApJ...660.1039K,2008A&A...478..311S}.   
% For a reliability check, we fit the PSF to nearby stars in Appendix \ref{sec:PSF-PSF}, confirming that the over/under-subtraction at the centre is not significant.
% Since there is no flux left other than the very centre, we conclude the QSO is consistent to be a PSF in the $z_r$ band as well. 
% We have computed a 1-$\sigma$ upper limit of the host brightness by randomly placing the 26-pix aperture on the random places of the image. 
%The upper-limit of the host galaxy is estimated to be $>25.34$ and $>24.22$ mag in  $i'$  and $z_r$ bands (See section \ref{SED}).
% $>$26.94 and $>25.69$ mag in  $i'$  and $z_r$ bands.

%\subsection{Radial Profile}

%\subsubsection{Host or Ly$\alpha$ blob?}
  Because the Ly$\alpha$ emission at z=6.43 is redshifted into the $z'$-band wavelength range, 
the extended structure in the $z'$-band could be an extended nebula of Ly$\alpha$ emission (Ly$\alpha$ blob) extensively searched at z$\sim$3 \citep{2004AJ....128..569M,2008ApJ...675.1076S}.  However, in this QSO image, the extended structure is also found in the $z_r$-band where the Ly$\alpha$ emission does not contaminate the flux. Therefore, it is more likely that at least some $z'$-flux stems from the host galaxy (continuum), and the $z'$-band flux is a mixture of host and its Ly$\alpha$ emission. We estimate the relative strength of the flux of the host (continuum) and  Ly$\alpha$ emission in detail in Section \ref{SED}.

\begin{figure}
\begin{center}
\includegraphics[scale=0.42]{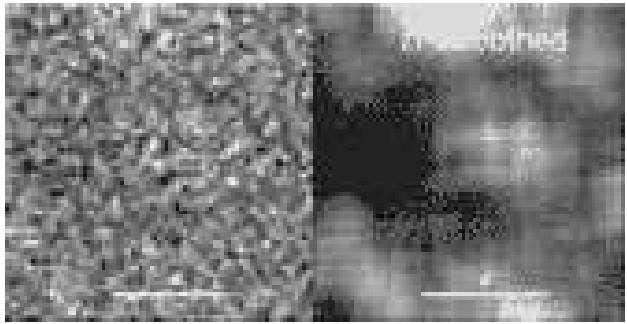}
\end{center}
\caption{
Both panels show residuals from the PSF subtraction in the $z_r$-band.
 The right panel is box-car smoothed with 10pix.
 The figures are north up, east  left.
}\label{fig:zr_smooth}
\end{figure}

\subsection{SED, stellar mass, SFR}\label{SED}

\begin{figure}
\begin{center}
\includegraphics[scale=0.65]{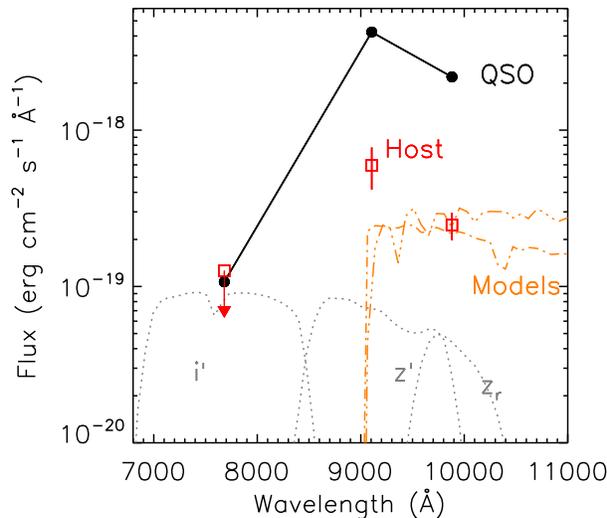}
\end{center}
\caption{
SEDs of QSO and its host galaxy. 
%The dash-dotted lines are SED models of 0.01, 0.1, 0.5 Gyrs of age with constant SFR.
Overplotted are SED models of constant SFR and delta starburst with 100 Myrs of age.
The host is not detected in $i'$-band, where 1$\sigma$ upper limit is shown.
}\label{fig:SED}
\end{figure}

In Fig.\ref{fig:SED}, we show the SED of the QSO (the black filled circles) and the host (the red rectangles). Since the host was not detected in the $i'$, 1 $\sigma$ upper limits are shown at the effective wavelength. We computed the 1 $\sigma$ of the sky noise by randomly placing an aperture of the same size on the blank region of the images, following the prescription by \citet{2002AJ....123...66Y,2004ApJ...611..660O}. 

 Overlaid are constant star-formation and delta burst models \citep{2003MNRAS.344.1000B} with the ages of 100Myrs scaled to the $z_r$-band flux. Assumed are the Salpeter initial mass function, solar metallicity  and no dust extinction. 
 Flux bluewards the Ly$\alpha$ emission is attenuated using the prescription given by \citet{2006ARA&A..44..415F}.
 These models are often used to compute colors of Lyman break galaxies at $z\sim$6.

 Comparing the colors of the host to the galaxy models, one immediately notices that the $z'$-band flux is not consistent with the flat continuum of the SED models, regardless of the star formation history. Perhaps this is because the host galaxy has some  Ly$\alpha$ emission in the $z'$-band.
 In the past, \citet{2008Msngr.131....7D} detected extended line emission in high redshift galaxies on scales of $\sim$4kpc.
 Such extended Ly$\alpha$ emission nebulae called Ly$\alpha$ blobs have been extensively searched  at $z\sim$3 \citep{2004AJ....128..569M,2008ApJ...675.1076S}.
 A QSO BR1202-0725 has a Ly$\alpha$ emitting companion, possibly ionized by the QSO \citep{1996ApJ...459L..53H,2004AJ....128.2704O}. 
 A typical FWHM of Ly$\alpha$ blobs is 300 km s$^{-1}$ \citep{2004AJ....128..569M,2008ApJ...675.1076S}.
  If the Ly$\alpha$ emission around the QSO is extended, it may be theoretically predicted infalling gas to form a galaxy, illuminated by a central QSO \citep{1988MNRAS.231P..91R}. If so, quantifying this will provide us with an important constraint on the formation of galaxies such as gas fraction, AGN geometry, and covering factors in the early Universe.

 Both of the SED models show an almost flat continuum slope at this small wavelength range between $z'$ and $z_r$-bands. Thus scaling from the $z_r$-bands flux and the SED models, in the $z'$-band, $\sim$2.5$\times$10$^{-19}$~erg~cm$^{-2}$~s$^{-1}$~\AA$^{-1}$ flux is estimated to be the light from the continuum of the host galaxy.  Subtracting this from the observed flux in the $z'$-band (6.0$\times$10$^{-19}$ erg cm$^{-2}$ s$^{-1}$ \AA$^{-1}$), $\sim$3.5$\times$10$^{-19}$ erg cm$^{-2}$ s$^{-1}$ \AA$^{-1}$ of flux is estimated to originate from Ly$\alpha$ emission of the host galaxy, i.e., in the $z'$-band, the continuum to Ly$\alpha$ flux ratio is 4:6. 

 If such flux came through the Ly$\alpha$ emission of 300 km s$^{-1}$,  then the  Ly$\alpha$ luminosity would be 1.6$\times 10^{42}$ erg s$^{-1}$, which are comparable luminosity to Ly$\alpha$ emitters at $z\sim$6 \citep{2006ApJ...648....7K,2008arXiv0807.4174O} and Ly$\alpha$ blobs at z=3.1\citep{2006ApJ...640L.123M}. 

 Based on the Ly$\alpha$ luminosity, we estimate the SFR(Ly$\alpha$) is  $>$1.6 $M_{\odot}$ yr$^{-1}$ \citep{1998ARA&A..36..189K}. 
 However, the SFR estimated here are lower limits because significant fraction of bluer side of the Ly$\alpha$ emission could be absorbed by the neutral hydrogen. 

%\subsection{SFR}
%At z=6.43, the absolute magnitude of the host to be $M_{1450}$=-22.88. 

We can also estimate star formation rate of the host galaxy based on the UV luminosity using \citet{1998ARA&A..36..189K}. Since the UV continuum of the host is relatively bright $M_{1450}$=-23.9, %($z_r$=24.3$\pm$0.2),
 we  obtain SFR$\sim$50 $M_{\odot}$ yr$^{-1}$. 
 It has been suggested that SFR derived from UV continuum is often larger than that from Ly$\alpha$, especially for Ly$\alpha$ emitters \citep{2004AJ....127..563H,2003AJ....126.2091A,2003PASJ...55L..17K}, as is the case here. 
Even though, this is quite a large number compared to that obtained through Ly$\alpha$ ($>$1.6 $M_{\odot}$ yr$^{-1}$), and thus, suggests the nature of the host galaxy; its UV continuum is brighter than typical Ly$\alpha$ emitters. 
 The host may be a massive galaxy that is decreasing its star formation activity.

 In the literature, \citet{2004ApJ...614..568J} reported 2-30  $M_{\odot}$ yr$^{-1}$ for z=1.8-2.6 QSO hosts. \citet{2008A&A...488..133V} measured 0.03-33 $M_{\odot}$ yr$^{-1}$ for QSO hosts at z=0.87 and z=2.75.
    The SFR estimated by UV light, however, is also subject to reddening uncertainties since a possible presence of dust has not been taken into account.

% Thus, we cannot rule out that some of the light in the $z'$-band came through the Ly$\alpha$ emission.
% However, it is also not plausible that the object is emission-only Ly$\alpha$ blobs; at $z\sim$3, Ly$\alpha$ blobs have much more irregular shapes, and the extended structures are hardly detected in the broad band \citep{2004AJ....128..569M,2008ApJ...675.1076S}. 
% Although it is not significant, the morphology of the QSO+host in the $z_{r}$ in the bottom left panle of Fig.\ref{fig:ip_subtraction} may have some similarities with that in $z'$-band in the left panel of Fig.\ref{fig:z_subtraction}. 
% Even in case of a giant Ly$\alpha$ emitter of L(Ly$\alpha$)=3.9$\pm0.2 \times 10^{43}$ erg s$^{-1}$, the extension in the broad band is $\sim$1'', which is much smaller than the  
%it is a possibility that the emission is through  Ly$\alpha$ and thus, the object is a Ly$\alpha$ blob at z=6.43.
  %However, it is rather unrealistic for an emission to boost the broad band flux by a factor of $\sim$4. 
%  Spatially-resolved spectroscopy and deeper photometry in the $z_r$-band would clarify the issue.

 With the lack of infrared photometry, it is difficult to constrain age/stellar mass through the SED.
 However, in the literature, a very young age of 5-100 Myrs for high redshift galaxies at 5$<z<$7 has been reported \citep{2005ApJ...618L...5E,2007ApJ...655..704L}.
 If we assume an age of 100 Myrs and scale models to $z_r$-band flux, we obtain a stellar mass of 1.1$\times 10^{10} M_{\odot}$ for the delta burst model, and 6.2$\times 10^{8} M_{\odot}$ for the constant SFR model. 
 Considering the age of the Universe is 840 Myrs, the host may be a forming massive galaxy, possibly co-evolving with the super massive black hole.
%By changing the assumed age within a range of 10-500 Myrs, the stellar mass changes by a factor of $\sim$5 with the constant SFR model. 

% With this caveat in mind, we could try to estimate the stellar mass of the host galaxy based on the $z$-band flux.
% Since the $z$-band is restframe UV light, the estimate heavily depends on the assumed age of the host. The range of the estimated stellar mass is 9.7$\times 10^6$-4.1$\times 10^8 M_{\odot}$ for 0.01-0.5 Gyrs of ages. Although the host is very luminous in the restframe UV with $M_{1450}$=-22.88, this exercise shows a moderately massive galaxy with extensive star formation can explain the flux. 

%Maybe galaxy has significant emission? Something is wrong with the zr mag? It has to be 3 sigma not 1 sigma.
%xxx

\subsection{S\'ersic  fit}

\begin{figure*}
\begin{center}
\includegraphics[scale=0.9]{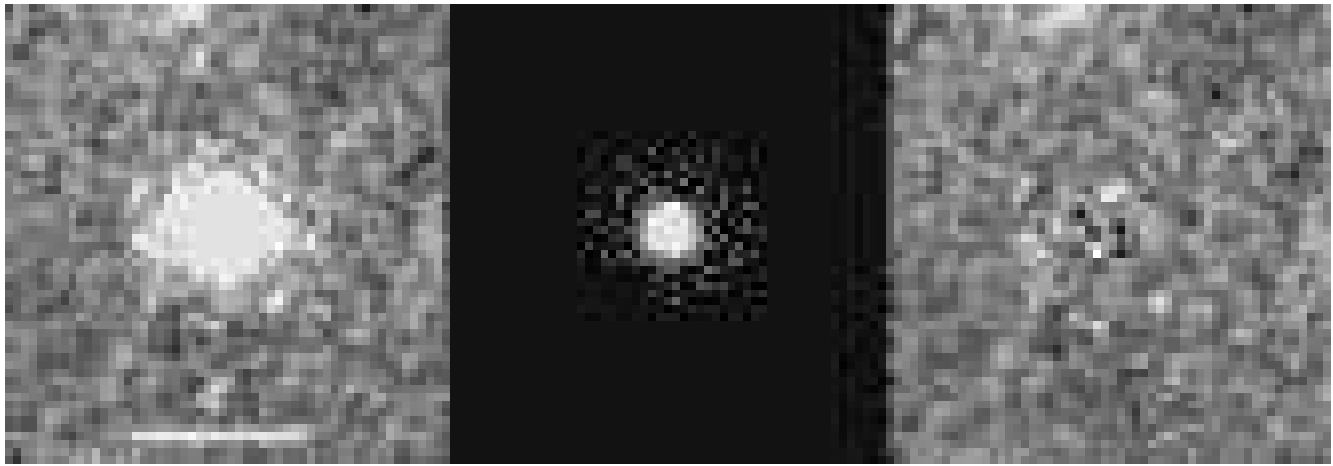}
\end{center}
\caption{
(left) The original $z'$-band image of the QSO.
(middle) the best fit PSF+s\'ersic model.
(right) residual from the model subtraction.
}\label{fig:z_ser_subtraction}
\end{figure*}

 Since we have a clear detection of extended structure in the $z'$-band,
 we attempted to fit the $z'$-band image with a PSF+s\'ersic+sky model using the {\ttfamily  galfit}. 
 The $z_r$-band image does not have enough signal-to-noise ratio to include S\'ersic profile in the fit.
 Results are shown in Fig.\ref{fig:z_ser_subtraction}. The middle panel of Fig.\ref{fig:z_ser_subtraction} shows the PSF+s\'ersic  (PSF convolved) model, which is subtracted from the original image shown in the left panel. 
  Some residual can be seen in the right panel, possibly implying an irregular morphology of the host, being consistent with  \citet{2006AJ....131..680H} who pointed out that high redshift QSO host galaxies are found to be irregular and highly disturbed.
 Measured s\'ersic  parameters are $R_e$=2.0''$\pm$0.1, axis ratio of b/a=0.74$\pm$0.03, and $n=0.019$, suggesting that the host possesses a flatter profile than the PSF, and even than de Vaucouleurs. 
 In the left panel of Fig.\ref{fig:radial_profile}, we show the PSF+S\'ersic model in the green  short-dashed line.
 The PSF+S\'ersic model in the short-dashed line is a good fit within 8 pixels of radius, although the model underestimates the profile significantly at $>$8 pixels of radius. 
  
 The extended structure has an effective radius of $R_e$=2.0''$\pm$ according to the s\'ersic  fit. Even in the image, one can easily recognize the light distribution  beyond 13 pixel (=2.6''). 
At the distance of z=6.43, 2.0'' corresponds to 11 kpc. 
% Therefore, the host certainly is a giant galaxy.
 As a reference, in the GOODS fields, a mean size of $z\sim5$ galaxies with 1-5$L^*$ is 0.3'' ($\sim$2 kpc) in restframe 1500\AA \citep{2004ApJ...600L.107F}.
 In the UDF, a mean size of $z\sim6$ galaxies with 0.1-1.0$L^*$ is 0.15'' ($\sim$1 kpc) in restframe 1600\AA \citep{2004ApJ...611L...1B}.
% It is surprising to find such a giant galaxy at z=6.43 when the age of the Universe is only 840 Myrs old.
 
On the size of QSO hosts in the literature, \citet{2001MNRAS.326.1533K}, who observed in $J$ and $Ks$, found a mean of 11.1 $\pm$ 5.7 kpc for their $z\approx1$ sample. For their $z\approx$ 1.5-2 sample they found a value of 9.78 $\pm$ 6.58 kpc for objects where it could be determined and upper limits ($<$10 kpc) for the remaining sources.
\citet{2007ApJ...660.1039K} found values of 6.7$\pm$1.7 kpc for their $z\approx$ 1-2 sample (observed in K and H). \citet{2008ApJ...673..694F}  found values between 2.6 kpc and 11.3 kpc for a sample of three high redshift QSOs ($z\approx$ 2-3) observed in Ks.
 \citet{2008A&A...488..133V} found  1.5-3.9 kpc  at z=0.9 and  1.5-3.9 kpc at z=2.8 in the FORS deep field.

However, direct interpretation of our results is difficult; in the last subsection we estimated that the $z'$-band flux is a mixture of 40\% continuum light and 60\%Ly$\alpha$ emission. Therefore, this extension of 11 kpc scale can be either extended Ly$\alpha$ emission or the extended stellar population of the host. Clear morphological separation of these two have to await high-resolution near-infrared imaging, or specially-resolved IFU spectroscopy in the red end of the optical CCD. A spectro-polarization is also important to constrain contribution from dust scattered light (Young et al. 2009, arxiv:0905.4102).

%この章は書き換えが必要！！！
%xxxx  Therefore, our detection of 11 kpc of host galaxies is at the largest end among the QSO host galaxies at; various redshift. 
% %Certainly it is surprising that the host is at z=6.43. 
%xxxx Only ours is at z=6.43. 
%xxxx It seems a luminous QSO at z=6.43 is already hosted by a giant galaxy as large as the host galaxies of QSOs at z$\sim$1.

%Guyon を引用して議論すること。ただし場所はここでない方がいいかもしれない。no-clear correlation reported in their paper. xxx
%For low redshift host galaxies clear correlations between QSOs and their host galaxies were found, with fainter QSOs residing in relatively faint late type galaxies and brighter QSOs residing in relatively bright elliptical galaxies.
% 

%

%13pix = 2.6''. Compare with 
%rest-frame UVでhalf-light radius~0.2''
%です。(Bouwens et al. 2004, Ferguson et al. 2004)と比較。

%IDL> print, zang(11,6.43)
%LUMDIST: H0: 70 Omega_m: 0.30 Lambda0 0.70 q0: -0.55 k:  0.00
%      2.00216

\section{Discussion}
\label{discussion}

%  \item Is this a foreground lensing galaxy?
In Section \ref{results}, we detected an extended light in the $z'$ and $z_r$-band, and suggested that this was the QSO host galaxy at z=6.43. However, it is always a possibility that the extended light could be from a foreground lensing galaxy, which also explains the bright magnitude of the QSO by lensing magnification.
 However, we believe it is not very likely in this case; using the 1$\sigma$ limit of $i'>$25.34, the $i'-z'$ color of the host is $>2.1$, which is very red and is hard to be explained unless the host has a break between $i'$ and $z'$. One possibility is a Balmer break at z=1.25. However, the color of the Balmer break is bluer, at approximately 0.64$<u-g<$1.99 \citep{1995PASP..107..945F}. The size, absolute magnitude, nuclear-to-host flux ratio all do not seem to be unreasonable for the structure to be at z=6.43.
 In addition, no significant flux is detected at 6500\AA$<\lambda<$8900\AA~ in the GMOS spectrum with one hour of exposure time \citep{2007AJ....134.2435W}. 
 Therefore, it is more plausible to consider the extended emission is from the host galaxy of the QSO at z=6.43.

% \item extended emission?  
% Since the host detection is only in the $z'$-band, we need to be careful that the extended light in the $z'$-band may stem from the Ly$\alpha$ emission instead of the continuum of the host galaxy.
% For example, Davies et al. (2008) detected extended line emission in high redshift galaxies on scales of $\sim$4kpc. \citet{2004AJ....128..569M} detected extended Ly$\alpha$ emission often extended to $>$10'' at z=3.1. 
%We might have detected something similar.
% However, the $z'$-band flux is 23.63 mag. If this is to be explained only with the  Ly$\alpha$ emission, unrealistically large luminosity of xxx is required.
 % However, the color of broad - narrow band colors of these Ly$\alpha$ blobs is $BV-NB\sim 1.5$. 
 
% \item 絶対等級 -22.88. QSO vs host luminosity との比較。
We have estimated the host luminosity to be $M_{1450}$=-23.9, i.e., the QSO to host flux ratio in restframe UV is 8.8. 
This is a higher ratio compared to the previous work. \citet{2005PASP..117.1250H} found nuclear-to-extended ratio of 2.3 in rest 2100\AA. \citet{2006ApJS..166...89G} report a flux ratio of 2.5 (note their pass-bands are in NIR).
 If the mass of the black hole is proportional to that of the host \citep{1998AJ....115.2285M,2003Ap&SS.284..565I}, and if the QSO emits at a fixed fraction of the Eddington luminosity, then one would expect a correlation between the luminosities of the host and QSO.  However, a number of reason can explain the deviation; the host may be more actively star-forming at earlier epoch; the Eddington ratio may be different at z=6.43; the amount of nuclear obscuration may be different.
\citet{2006AJ....131..680H} reported a higher QSO-to-host flux ratio for more luminous QSOs among their sample.
% Our detection of high flux ratio might imply the relation does not hold at z=6.43. 
Since this is only one example based on the restframe UV photometry, it is important to obtain infrared photometry to accurately measure the stellar mass/ luminosity ratio of more hosts. High resolution spectroscopy is also important to examine $M-\sigma$ relation at z$>$6.

% \begin{itemize}
%  \item SED fitting. mass estimate.
%  \item Guyon, host mag vs QSO magの議論。 
%  \item Comparison to the previous work. 
%  \item もしかして、ｚｒバンドで中央に受かっているのか？
%  \item host type, age, SFR, BH mass

%  \item radial profileを書くこと。irafでも良いので。これは重要！
% \end{itemize}

%\subsection{Comparison to previous work}

%In this section we compare our results to those of other authors and discuss the results regarding the link between galaxy evolution and QSO activity and the evolution of supermassive black holes.

%The good correlation of QSO and host luminosity leads to the question if QSO activity and host galaxy formation are directly linked. i.e. are both phenomena triggered by the same event or does one of the events trigger the other. 

% In the literature, QSO host galaxies have been studied mainly on nearby, bright QSOs. 
Comparing the morphology of the host to those in the literature,
\citet{2003AJ....125.1053H} resolved the host galaxies of 4 high redshift QSOs ($z\sim4.7$) and found heavily disturbed galaxies, partially showing signs of ongoing merging. 
\citet{2006ApJS..166...89G} reported that 30\% of PG QSO host galaxies showed obvious sign of disturbances, with strongly disturbed hosts favoring more luminous QSOs.
 In this work, there were some residuals after the host was fit with a s\'ersic profile in the right panel of Fig.\ref{fig:z_ser_subtraction}, suggesting a possible non-smooth morphology.
Further morphological investigation, however, is difficult due to the limited depth and spatial resolution of our ground-based data. It is necessary to take high-resolution near-infrared AO or HST data to investigate the morphology of the host galaxy.

\section{Summary}

 We have detected an extended structure around a QSO at z=6.43 in the $z'$ and $z_r$-band images of the Subaru/Suprime-Cam. The QSO is the highest redshift QSO currently known.
 After a careful PSF subtraction, the structure is detected at a 16 $\sigma$ level in $z'$-band and at a 3$\sigma$ level in the $z_r$-band.
 In the $z'$-band, the radial profile of the QSO+host shows clear excess at $>4$ pix of radius over the carefully constructed PSF. There is a marginal excess in the $z_r$-band profile at 6-10 pix of radius as well. 
 Through the SED modeling, we estimate 40\% of the PSF-subtracted $z'$-band light is from the host (continuum) and 60\% is from Ly$\alpha$ emission of the galaxy.

% Therefore, we claim a detection of the QSO host galaxy at z=6.43. 
% This is the highest redshift QSO host as a date of writing.

 The host is a bright, large galaxy with  $M_{1450}$=-23.9. Its extended structure has an effective radius of $>$11 kpc. 
 The lower limit on the  SFR(Ly$\alpha$) is $>$1.6 $M_{\odot}$ yr$^{-1}$, the stellar mass of the host ranges from  6.2$\times 10^8$ to 1.1$\times 10^{10} M_{\odot}$ depending on the star-formation history when 100 Myrs of age is assumed.
 Therefore, the host is a large, star-forming galaxy, with a moderate amount of stellar mass.

% The host is on the host-QSO relation?
% stellar-mass?

% This object may provide us with an important observational constraints how galaxies and QSOs form in the early Universe.  
 These results present an important example that indeed a super massive black hole resides in a large galaxy even in the early epoch when the Universe is $\sim$840 Myr old. 
  Quantifying the host mass and fraction of ionized gas in more detail will provide us with an important constraint on how galaxies and AGN form in the early Universe, such as gas fraction, AGN geometry, and covering factors.

\section*{Acknowledgments}

We thank the anonymous referee for many insightful comments, which significantly improved the paper.
%We are grateful to Hisanori Furusawa for valuable help during the observation.
%We are grateful to T.Hattori for valuable help during the observation.
%We thank Youichi Ohyama for useful discussions. %helpful suggestions. t
%and Dr. T.Hattori for friendly help during the
%observation. 

%Dr H. C. Bhatt for a critical reading of the original version of the
%paper and an anonymous referee for very useful comments that improved
%the presentation of the paper.

T.G. acknowledges financial
support from the Japan Society for the Promotion of
Science (JSPS) through JSPS Research Fellowships for Young
Scientists.

This work is supported in part with the research fund for students (2008) of the Department of Astronomical Science, the Graduate University for Advanced Studies, Japan.

This work is also supported with a Grant-in-Aid for Scientific Research on Priority Areas (18072003) from the Ministry of Education, Science, Culture, and Sports of Japan (MEXT). 

% Use of the UH 2.2-m telescope for the observations is supported by NAOJ.
% The research was financially supported by the Sasakawa Scientific Research Grant from The Japan Science Society.
% This research was partially supported by the Japan Society for the Promotion of Science through Grant-in-Aid for Scientific Research 18840047.
%
The authors wish to recognize and acknowledge the very significant cultural role and reverence that the summit of Mauna Kea has always had within the indigenous Hawaiian community.  We are most fortunate to have the opportunity to conduct observations from this sacred mountain.
%
%    Funding for the creation and distribution of the SDSS Archive has been provided by the Alfred P. Sloan Foundation, the Participating Institutions, the National Aeronautics and Space Administration, the National Science Foundation, the U.S. Department of Energy, the Japanese Monbukagakusho, and the Max Planck Society. The SDSS Web site is http://www.sdss.org/.
%
%    The SDSS is managed by the Astrophysical Research Consortium (ARC) for the Participating Institutions. The Participating Institutions are The University of Chicago, Fermilab, the Institute for Advanced Study, the Japan Participation Group, The Johns Hopkins University, Los Alamos National Laboratory, the Max-Planck-Institute for Astronomy (MPIA), the Max-Planck-Institute for Astrophysics (MPA), New Mexico State University, University of Pittsburgh, Princeton University, the United States Naval Observatory, and the University of Washington.

%\clearpage

\appendix

\section[]{PSF subtraction test}
\label{sec:PSF-PSF}

\begin{figure}
\begin{center}
\includegraphics[scale=0.33]{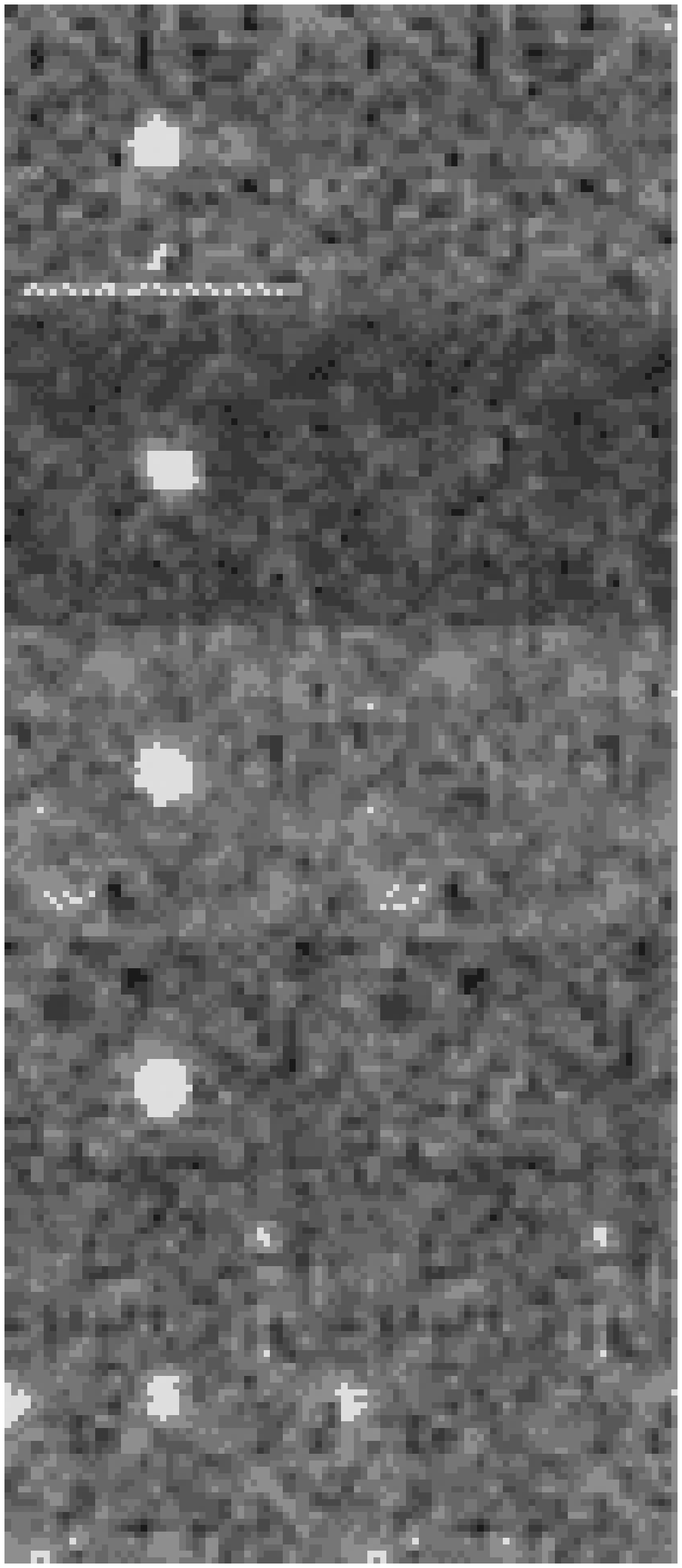}
\includegraphics[scale=0.33]{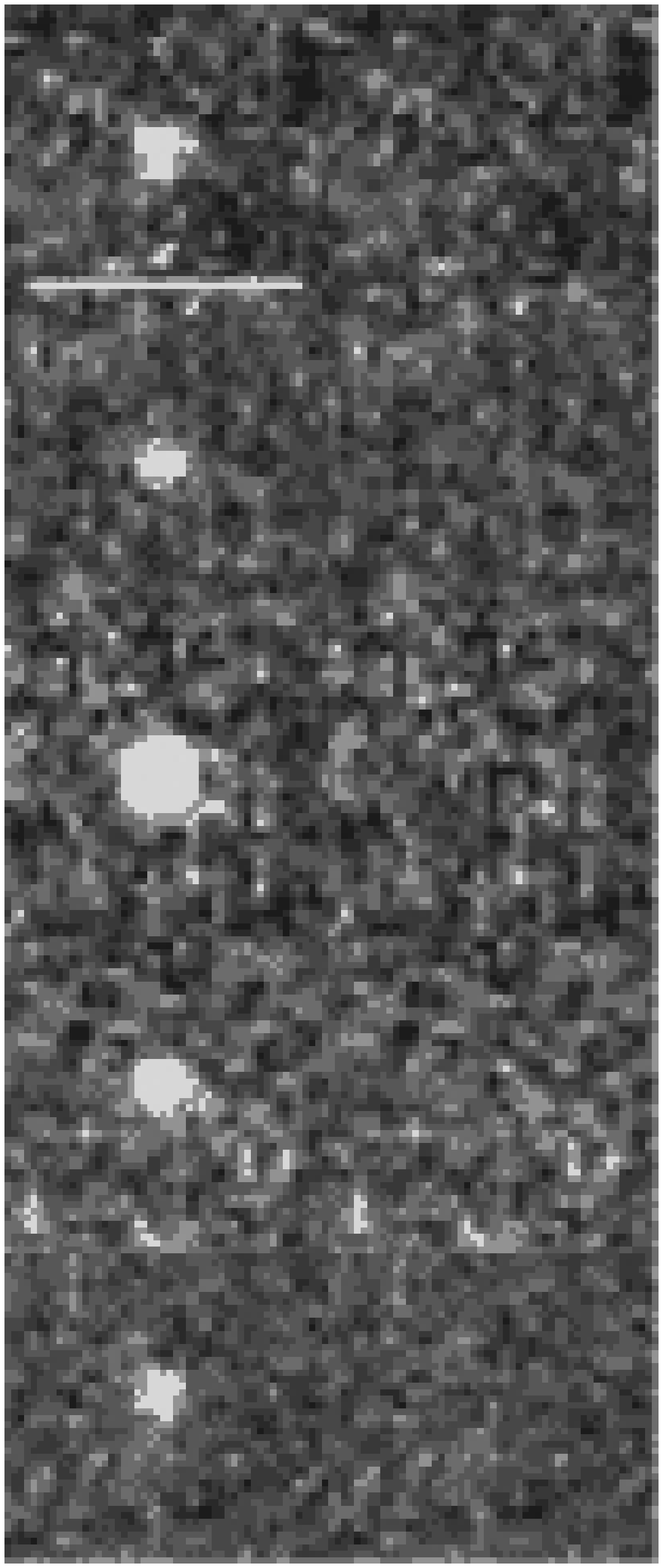}
\end{center}
\caption{
Reliability of the PSF subtraction is tested by fitting the constructed PSF from 5 nearby stars, in $i'$ and $z'$-bands.
In each band, the left panels show actual stellar images.
%(middle) PSF constracted using nearby stars
The right panels show residuals from the PSF-subtraction. 
}\label{fig:subtraction}
\end{figure}

\begin{figure}
\begin{center}
\includegraphics[scale=0.5]{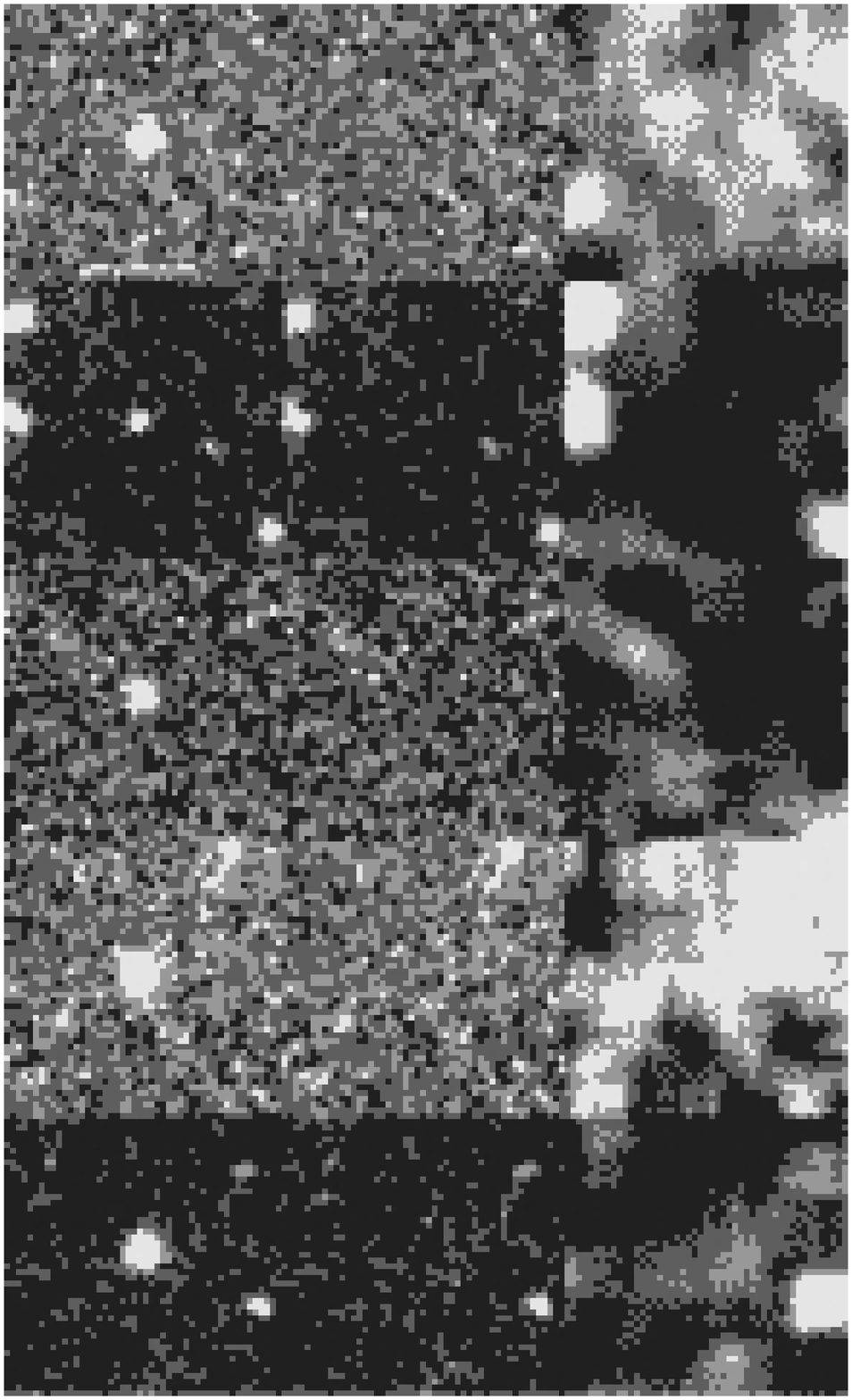}
\end{center}
\caption{
Reliability of the PSF subtraction is tested by fitting the constructed PSF to 5 actual stars in the $z_r$-bands.
The left panels show actual stellar images.
%(middle) PSF constracted using nearby stars
The middle panels show residuals from the PSF-subtraction. 
The right panels are 10-pixel box-car smoothed images of the residual.
}\label{fig:zr_subtraction}
\end{figure}

Fitting the constructed PSF (Sec. \ref{sec:psf}) to the observed stars around QSO provides a sanity check of the PSF subtraction.
In this exercise, we first select stars within 200'' around the QSO as objects with source extractor's stellarity of $\ge 0.97$. We have specifically picked stars as bright as, but not brighter than the QSO itself to avoid a possible non-linearity effect at the central pixels of bright stars. %正しいか？
These are the stars used to construct a composite PSF in Section \ref{results}. In Fig.\ref{fig:subtraction}, we show 5 images of stars (left panels) and PSF-subtracted images (right panels) in $i'$(left) and $z'$(right)-bands.  Fig.\ref{fig:zr_subtraction} shows the same PSF subtraction for $z_r$-band, but the rightmost panels show 10-pixel box-car smoothed images of residuals (the middle panels) in the $z_r$-band.

None of the residuals in Fig.\ref{fig:subtraction} shows an extended structure such as seen in the $z'$ in the middle right panel of Fig.\ref{fig:all_subtraction}.
 In the $z_r$-band, no obvious residuals are seen even after the images are box-car smoothed (c.f. Fig.\ref{fig:zr_smooth}).
 These sanity checks suggest that the extended structures found in Figs. \ref{fig:all_subtraction} and \ref{fig:zr_smooth} are not  artifacts from the PSF subtraction.

% It is immediately noticed that some images have remaining flux at the centre after the subtraction. In the $z'$-band, some over-subtraction can be seen in the bottom three panels. These are results of incomplete pixeralization and it rather happens often especially at the centre where S/N of the fit is poorer. Therefore, neither the over-subtraction seen in the right panel of Fig.\ref{fig:z_subtraction} nor under-subtraction in the right panel of Fig.\ref{fig:ip_subtraction} is significant.

%\begin{figure*}
%\begin{center}
%\includegraphics[scale=0.9]{/home/tomo/work/galfit-example/utsumi_QSO/081209_galfit_QSOhost.ps} %z1psf
%\includegraphics[scale=0.9]{/home/tomo/work/galfit-example/utsumi_QSO/081209_galfit_QSOhost0.ps} %z2psf
%\end{center}
%\caption{
%(left) original $z'$-band image
%(middle) PSF constracted using nearby stars
%(right) PSF subtracted image. 
%}\label{fig:subtraction}
%\end{figure*}

%\begin{figure*}
%\begin{center}
%%\includegraphics[scale=0.9]{/home/tomo/work/galfit-example/utsumi_QSO/081210_galfit_QSOhost.ps}
%\includegraphics[scale=0.9]{/home/tomo/work/galfit-example/utsumi_QSO/081209_galfit_out_psf-psf.ps}
%\end{center}
%\caption{
%(left) original $z'$-band image of a star
%(middle) PSF model
%(right) residual from model subtraction. 
%}\label{fig:psf_psf}
%\end{figure*}

%\bsp

\label{lastpage}

\end{document}